\documentclass[prb,twocolumn,showpacs,preprintnumbers,superscriptaddress,floatfix]{revtex4}
\usepackage{amsfonts}
\usepackage{amsmath}
\usepackage{amssymb}
\usepackage{graphicx}
\usepackage[rightcaption]{sidecap}

\begin{document}

\preprint{}
\title[Electronic Raman Scattering from a Multiband Superconductor]{Theory
of Raman scattering from Leggett's collective mode in a multiband
superconductor: Application to MgB$_{2}$}
\author{M. V. Klein}
\affiliation{Department of Physics and Frederick Seitz Materials Research Laboratory,
University of Illinois at Urbana-Champaign, Urbana IL 61801}
\author{}
\affiliation{}
\author{}
\affiliation{}

\begin{abstract}
In 1966 Leggett used a two-band superconductor to show that a new collective
mode could exist at low temperatures, corresponding to a counter-flow of the
superconducting condensates in each band. Here, the theory of electronic
Raman scattering in a superconductor by Klein and Dierker (1984) is extended
to a multiband superconductor. Raman scattering creates particle/hole pairs.
In the relevant $A_{1g}$\ symmetry, the attraction that produces pairing
necessarily couples excitations of superconducting pairs to these p/h
excitations. In the Appendix it is shown that for zero wave vector transfer $%
q$ this coupling modifies the Raman response and makes the long-range
Coulomb correction null. The 2-band result is applied to MgB$_{2}$ where
this coupling activates Leggett's collective mode. His simple limiting case
is obtained when the interband attractive potential is decreased to a value
well below that given by LDA theory. The peak from Leggett's mode is studied
as the potential is increased through the theoretical value: With realistic
MgB$_{2}$\ parameters, the peak broadens through decay into the continuum
above the smaller ($\pi $ band) superconducting gap. Finite $q$ effects are
also taken into account, yielding a Raman peak that agrees well in energy
with the experimental result by Blumberg \textit{et el.} (2007). This
approach is also applied to the $q=0$, 2-band model of the Fe-pnictides
considered by Chubukov \textit{et al. }(2009).
\end{abstract}

\volumenumber{number}
\issuenumber{number}
\eid{identifier}
\date[Date text]{date}
\received[Received text]{date}
\revised[Revised text]{date}
\accepted[Accepted text]{date}
\published[Published text]{date}
\startpage{1}
\endpage{2}
\maketitle


\section{Introduction}

In a superconductor, a longitudinal density oscillation corresponds to an
oscillation of the phase of the superconducting order parameter. As shown by
Bogolyubov \textit{et el. }\cite{Bogo} and Anderson, \cite{PWA} this
oscillation would amount to a collective mode with zero energy
at zero wavevector in the absence of the long-range Coulomb interaction
between the conduction electrons. The presence of this interaction, however,
causes the condensate density oscillation to become the electronic plasma
oscillation, typically with an energy of 5-10 eV. Leggett showed that for a
simple model of a two-band superconductor it could be possible for the
superconducting condensates in each band to oscillate longitudinally
relative to one another, equivalent to an oscillation in the relative phase
of the two condensates, with the net displacement of the total condensate
being zero. \cite{Tony} The frequency of this mode and its dispersion with wavevector $q$ would depend substantially on the effective interband pairing potential. If this potential were sufficiently small, but not zero, the small $q$ mode frequency would be less than the smaller of the superconducting gaps $2\Delta $, for the two bands. In the low temperature limit this mode would become a true collective
mode with an infinite lifetime.

There is good evidence that MgB$_{2}$ is a multiband superconductor. Whereas
theory suggests that there are 2 bands primarily of $2p\sigma $ character
and 2 bands of $2p\pi $ character, \cite{Liu} most experimental
superconducting properties can be described as though there were one $\sigma 
$ band and one $\pi $ band. \cite{bandstructure} The $\sigma $ band is
nearly two-dimensional. Its barrel-shaped Fermi surface is almost
cylindrically-symmetric about the $c$ axis and the $c$ axis component of the
Fermi velocity is very small. On the other hand, the $\pi $ band Fermi
surface is three dimensional. On it the root mean square Fermi velocity
components are large and nearly isotropic.

Electronic Raman scattering measurements in $E_{2g}$ symmetry $(xx-yy,$ or $%
xy)$ show two features due to superconductivity: a relatively sharp peak at $%
2\Delta _{\sigma }=13.5$ meV associated with the $\sigma $ band and an onset
without a peak at $2\Delta _{\pi }=4.6$ meV associated with the $\pi $ band. 
\cite{Blumbergphonons} As shown by Blumberg \textit{et el.}, the situation
is different for Raman in the fully-symmetric $A_{1g}$ symmetry $(xx+yy)$. 
\cite{Blumbergelectrons} Apart from two-phonon features, an electronic Raman
peak was found at 9.4 meV, midway between the onset of the $\pi $ band
continuum at $2\Delta _{\pi }$ and the $\sigma $ band peak at $2\Delta
_{\sigma }$. Because it can decay into the $\pi $ band continuum, the mode
signified by this peak cannot be a true collective mode. We will show that
it represents a resonance mode that follows by continuity from Leggett's
collective mode when allowance is made for (1) a vertex correction that is
related to the mechanism that allows electronic Raman scattering to couple to
it, (2) damping due to decay into the $\pi $ band continuum, (3) necessary
kinematic corrections that depend upon $\boldsymbol{q\cdot v}_{F},$ where $%
\boldsymbol{q}$ is the transferred wavevector and $\boldsymbol{v}_{F}$ is
the Fermi velocity vector, (4) integration over values of $q$ that are
ill-defined due to optical absorption, all using a realistic set of physical
parameters that apply to MgB$_{2}$.

The present work is based on a paper by Klein and Dierker (K-D) \cite{KD}
that uses a Green's function approach to describe the electronic Raman
response function as resulting from Kawabata's 4-vertex function \cite%
{Kawabata} The latter may be collapsed into a 2-vertex polarization
\lq bubble.\rq\, The bare vertex $\gamma _{k}$ comes from the electron-photon
interaction and is the sum of two terms: the $A^{2}$ term, where $%
\boldsymbol{A}$ is the vector potential, couples to the electron density
operator in first order, and the $\boldsymbol{p\cdot A}$ term couples in
second order, ($\boldsymbol{p}$ is the electron momentum operator) to a
density-like quantity through a virtual intermediate state. The sum of these
terms, $\gamma _{k},$ gives the amplitude for excitation from near the FS to
near the FS of an electron from wavevector $\boldsymbol{k}$ to wavevector $%
\boldsymbol{k}+\boldsymbol{q}$ while the incident laser $(L)$ photon is
transformed into the scattered $(S)$ photon with respective wavevector and
energy transfers $\boldsymbol{q}=\boldsymbol{k}_{L}-\boldsymbol{k}_{S}$ and $%
\omega =\omega _{L}-\omega _{S}$. For small values of $q$ relative to the
Fermi wavevector, the vertex $\gamma _{k}$ does not depend on $\boldsymbol{q}
$. Due to the $\boldsymbol{p\cdot A}$ term, $\gamma _{k}$ will depend on $%
\omega _{L}$ when the excitation energy to the intermediate state comes into
resonance with $\omega _{L}$. Such resonance Raman effects are probably not
relevant for the Blumberg \textit{et el. }data taken with 1.65 eV laser
light: Optical data \cite{optics} and band structure calculations \cite{Liu}
show that interband optically-allowed transitions to and from the Fermi
surface in MgB$_{2}$ occur at energies well above 1.65 eV. The \lq mass
approximation\rq\, for the relevant $A_{1g}$ symmetry may then be applied. It
says that the electronic Raman vertex $\gamma _{k}$ is $r_{o},$ the
classical radius of the electron, multiplied by $(\mu _{xx}(k)+\mu _{yy}(k))$%
, where $\boldsymbol{\mu }(k)$ is the inverse effective mass tensor at
wavevector $k$ near the FS. The holelike $\sigma $ band and the electronlike 
$\pi $ band possess $\mu _{xx}+\mu _{yy}$ values that have opposite signs.

To allow for the possibility of electronic resonance(s) with intermediate
state(s), we assume in our derivation that $\gamma _{k}$ is complex due to
the appearance of an imaginary term in the energy denominator(s)
proportional to the inverse lifetime of the intermediate state(s).

\section{Model and its Equations}

To carry out the calculation, we will set up the problem in a more general
fashion and then make the assumption that $\gamma _{k}$ is a positive
constant for $k$ in the $\pi $ band and a negative constant for $k$ in the $%
\sigma $ band. These bare vertices create a positive charge-density
(particle/hole or p/h) excitation of wave-vector $\boldsymbol{q}$ in the $%
\pi $ band and a negative charge-density excitation of wave-vector $%
\boldsymbol{q}$ in the $\sigma $ band. In the Nambu formalism \cite{Nambu}
used by K-D, these p/h excitations are associated with the $\tau _{3}$ Nambu
matrix. A generalized vertex equation necessarily couples the (p/h) $\tau
_{3}$ channel with the particle/particle (p/p) $\tau _{2}$ (phase) channel
that describes excitations of the condensate. This is the first of four
final state corrections to be consider in this paper, the \lq pairing
correction.\rq\, The opposite signs of the bare $\gamma _{k}$ in the p/h channel
produce opposite forcing terms in the p/p-phase channel for the 2 bands and
thus enable coupling to Leggett's collective mode. The dynamics of the mode
are altered because of the second final state correction that occurs in the
p/h channel, the \lq vertex correction.\rq

We begin by quoting key results from Sections III.2 and III.3 of K-D,
somewhat rewritten to fit our present needs. We assume that $\boldsymbol{k}$%
, hereafter denoted by $k$, is a wave vector near the Fermi surface,
referring to one or more bands, and consider a $k$-dependent superconducting
gap $\Delta _{k}$, normal state energy minus the Fermi energy $\varepsilon
_{k}$, and Fermi velocity $\boldsymbol{v}_{k}=\boldsymbol{\nabla }%
_{k}\varepsilon _{k}$ \ Define%
\begin{equation}
\beta _{k}\equiv \sqrt{\frac{\omega ^{2}-(\boldsymbol{q}\cdot \boldsymbol{v}%
_{k})^{2}}{4\Delta _{k}^{2}}}\text{; \ }f_{k}(\omega ,\boldsymbol{q})\equiv 
\frac{\arcsin \beta _{k}}{\beta _{k}\sqrt{1-\beta _{k}^{2}}};
\label{Def beta, f}
\end{equation}%
\begin{eqnarray}
p_{k}(\omega ,\boldsymbol{q}) &\equiv &f_{k}(\omega ,\boldsymbol{q})\left[
1-\left( \frac{\boldsymbol{q}\cdot \boldsymbol{v}_{k}}{\omega }\right) ^{2}%
\right] \text{;}  \label{Def p} \\
r_{k}(\omega ,\boldsymbol{q}) &\equiv &\frac{f_{k}(\omega ,\boldsymbol{q}%
)\omega ^{2}-(\boldsymbol{q}\cdot \boldsymbol{v}_{k})^{2}}{\omega ^{2}-(%
\boldsymbol{q}\cdot \boldsymbol{v}_{k})^{2}}  \label{Def r}
\end{eqnarray}%
\allowbreak

The correct branches of the square root and arcsin functions are obtained by
adding an infinitesimal positive imaginary part to the frequency variable $%
\omega ,$ which is itself assumed to be positive, and by assuming that
branch cuts are along the negative real axis of the arguments of the two
functions. Note that when $\boldsymbol{q=}0$, $p_{k}$\ and \ $r_{k}$ equal $%
f_{k}$, but when $\boldsymbol{q\neq }0$, they no longer equal $f_{k}$.

Assume zero temperature and that a non-retarded effective potential $%
D_{k,k^{\prime }}^{(p)}$ (pairing correction) acts between a p/p pair at $k$
and one at $k^{\prime }$ and that a potential $D_{k,k^{\prime }}^{(v)}$
(vertex correction) acts between a p/h pair at $k$ and one at $k^{\prime }$.
We obtain the two vertex equations, rearranged from K-D Eqs. (10a,b), that
connect the \lq dressed\rq\, vertices $\Gamma _{k}^{\left( 3\right) }$ and $\Gamma
_{k}^{\left( 2\right) }$ for the p/h and p/p-phase channels, respectively
with the bare p/h vertex $\gamma _{k}$:

\begin{equation}
\Gamma _{k}^{\left( 3\right) }=\gamma _{k}+\int dS_{k^{\prime
}}D_{k,k^{\prime }}^{(v)}\frac{i\omega }{2\Delta _{k^{\prime }}}\left[
f_{k^{\prime }}\Gamma _{k^{\prime }}^{\left( 2\right) }-\frac{2i\Delta
_{k^{\prime }}}{\omega }r_{k^{\prime }}\Gamma _{k^{\prime }}^{\left(
3\right) }\right] ,  \label{Vert eq gamma 3}
\end{equation}

\begin{eqnarray}
\Gamma _{k}^{\left( 2\right) } &=&-\int dS_{k^{\prime }}D_{k,k^{\prime
}}^{(p)}L_{k^{\prime }}\Gamma _{k^{\prime }}^{\left( 2\right) }-\int
dS_{k^{\prime }}D_{k,k^{\prime }}^{(p)}\frac{\omega ^{2}}{4\Delta
_{k^{\prime }}^{2}}  \notag \\
&&\mathsf{x}\left[ p_{k^{\prime }}\Gamma _{k^{\prime }}^{\left( 2\right) }-%
\frac{2i\Delta _{k^{\prime }}}{\omega }f_{k^{\prime }}\Gamma _{k^{\prime
}}^{\left( 3\right) }\right] .  \label{Vert eq gamma2}
\end{eqnarray}%
\newline

Here $dS_{k}=(2\pi )^{-3}d^{3}k\delta (\varepsilon _{k})$ and the Dirac
delta function $\delta (\varepsilon _{k})$ forces the $k$ integral to be
over the Fermi surface. $L_{k}\equiv \sinh ^{-1}(\omega _{k}/\Delta _{k})$
is a log-like term involving a cut-off energy $\omega _{k}$\ and the gap. It
also occurs in the BCS-like equation for the gap, K-D Eq. (18):

\begin{equation}
\Delta _{k}=-\int dS_{k^{\prime }}D_{k,k^{\prime }}^{(p)}L_{k^{\prime
}}\Delta _{k^{\prime }}  \label{gap eqn}
\end{equation}

Also needed is the equation for the polarization bubble, or Raman response
function, rearranged from K-D Eq.(10c):%
\begin{equation}
B_{\gamma ^{\prime },\Gamma \gamma }(\omega ,\boldsymbol{q})=-2\int
dS_{k}\gamma _{k}^{\prime }\frac{i\omega }{2\Delta _{k}}\left[ f_{k}\Gamma
_{k}^{\left( 2\right) }-\frac{2i\Delta _{k}}{\omega }r_{k}\Gamma
_{k}^{\left( 3\right) }\right] .  \label{Bubble}
\end{equation}

Here the notation $B_{\gamma ^{\prime },\Gamma \gamma }(\omega ,\boldsymbol{q%
})$\ means that the bare vertices are $\gamma _{k}$\ and \ $\gamma
_{k}^{\prime }$, with \ $\gamma _{k}$\ dressed by vertex corrections. The
argument $(\omega ,\boldsymbol{q})$ refers the fact that $f_{k}$, $p_{k}$, $%
r_{k}$, and $\Gamma _{k}^{\left( 2,3\right) }$ depend on the same argument,
omitted here for simplicity. The factor of $2$ outside the integral is from
a summation over spin inherent in the Nambu formalism.

In general, vertex corrections result from an infinite sum of ladder
diagrams. This has the consequence that the result is symmetric in the two
vertices. Thus\cite{sym}%
\begin{equation}
B_{\gamma ^{\prime },\Gamma \gamma }(\omega ,\boldsymbol{q})=B_{\gamma
,\Gamma \gamma ^{\prime }}(\omega ,\boldsymbol{q})\text{.}  \label{symmetry}
\end{equation}

Next we make the third final state correction by correcting the response
function for the long-range Coulomb interaction, K-D Eq. (17), to obtain the
screened response function

\begin{subequations}
\begin{align}
B_{\gamma ^{\ast },\Lambda \gamma }(\omega ,\boldsymbol{q})& =B_{\gamma
^{\ast },\Gamma \gamma }(\omega ,\boldsymbol{q})+\frac{B_{\gamma ^{\ast
},\Gamma 1}(\omega ,\boldsymbol{q})V_{q}B_{1,\Gamma \gamma }(\omega ,%
\boldsymbol{q})}{1-V_{q}B_{1,\Gamma 1}(\omega ,\boldsymbol{q})},
\label{screened bubble} \\
\text{and with }V_{q}& \equiv \frac{4\pi e^{2}}{q^{2}v_{c}}\text{.}
\label{defn Vq}
\end{align}%
Here $v_{c}$ is the volume of a unit cell. For values of $q$ that are small
compared to the inverse lattice constant 
\end{subequations}
\begin{equation}
B_{\gamma ^{\ast },\Lambda \gamma }(\omega ,\boldsymbol{q})\rightarrow
B_{\gamma ^{\ast },\Gamma \gamma }(\omega ,\boldsymbol{q})-\frac{B_{\gamma
^{\ast },\Gamma 1}(\omega ,\boldsymbol{q})B_{1,\Gamma \gamma }(\omega ,%
\boldsymbol{q})}{B_{1,\Gamma 1}(\omega ,\boldsymbol{q})}.
\label{screening corr}
\end{equation}

The appearance of $1$ on the right side of these equations signifies that
the relevant bare vertex is unity, 1, for all values of $k$. On the right
side of Eq.(\ref{screening corr}) the second term, the so-called \lq screening
correction,\rq\, projects off, in a $\boldsymbol{q}$ and $\omega $ dependent
fashion, from the bare vertex $\gamma _{k}$ that portion parallel to $1$.
When $\gamma _{k}\propto 1$, the projection cancels the first term, and the
screened response function is zero.

\section{Dealing with the spread in wave vector due to optical absorption}

The photon wavevectors are proportional to the complex index of refraction.
Thus, for Raman scattering from metals, the wavevector transfer is a complex
quantity $q=q^{\prime }+iq^{\prime \prime }$. The experiments of Blumberg $%
et $ $al.\cite{Blumbergelectrons}$ were such that $\boldsymbol{q}$\ was
parallel to the hexagonal symmetry axis. If $z$ denotes the distance along
the axis measured from the sample surface, then the perturbation that drives
the Raman response has spatial dependence $e^{i(q^{\prime }+iq^{\prime
\prime })z}$. This has a Fourier coefficient $(i/[\sqrt{2\pi }(q^{\prime
}+iq^{\prime \prime }-q)].$ The optical penetration depth, $(2q^{\prime
\prime })^{-1}$, is the order of 1/10th the wavelength of light and thus
much larger than a lattice constant. Therefore, Raman is essentially a bulk
probe, and excitations with different wavevectors $q$ are independent, that
is, non-interfering. The physical screened response function is obtained by
integrating $B_{\Lambda (\gamma ),\gamma ^{\ast }}(\omega ,q)|1/[\sqrt{2\pi }%
(q^{\prime }+iq^{\prime \prime }-q)]|^{2}$\ over $q$ from $-\infty $ to $%
\infty $. The Raman intensity $I(\omega )$\ is minus the imaginary part of
the result. Using the fact that $B_{\gamma ^{\ast },\Lambda _{\gamma
}}(\omega ,q)$ is even in $q$, we obtain a result equivalent to K-D's Eq.
(4a):

\begin{equation}
I(\omega )=-\int_{0}^{\infty }\frac{\textrm{Im}[B_{\gamma ^{\ast },\Lambda
\gamma }(\omega +i0^{+},q)](q^{\prime 2}+q^{\prime \prime 2}+q^{2})dq}{\pi %
\left[ \left( q^{\prime 2}+q^{\prime \prime 2}+q^{2}\right)
^{2}-4q^{2}q^{\prime 2}\right] }  \label{q-integrated int}
\end{equation}

\section{Application to the Multi-Band Case: Simplifying Assumptions}

We introduce $j$ as a band index and $k_{j}$ to denote wavevector values in
band $j$. The superconducting gap $\Delta _{j}$ and the magnitude of the
Fermi velocity $v_{j}$ are assumed to be constants in each band. Let $u$
denote the cosine of the angle between the direction of $\boldsymbol{q}$ and
the direction of the Fermi velocity $\boldsymbol{v}_{k_{j}}$ at $k_{j}$. The
quantities appearing in Eqs.(\ref{Def beta, f}-\ref{Def r}) will depend on
the discrete variable $j$ and the continuous variable $u$:%
\begin{eqnarray}
\beta _{k} &\rightarrow &\beta _{j}(\omega ,q,u)\equiv \sqrt{\frac{\omega
^{2}-(qv_{j}u\ )^{2}}{4\Delta _{j}^{2}}}  \label{multiband beta} \\
f_{k} &\rightarrow &f_{j}(\omega ,q,u)\equiv \frac{\arcsin \beta _{j}(\omega
,q,u)}{\beta _{j}(\omega ,q,u)\sqrt{1-\beta _{j}(\omega ,q,u)^{2}}}
\label{multiband f}
\end{eqnarray}%
\begin{eqnarray}
p_{k} &\rightarrow &p_{j}(\omega ,q,u)\equiv f_{j}(\omega ,q,u)\left[
1-\left( \frac{qv_{j}u}{\omega }\right) ^{2}\right]  \label{multiband p} \\
r_{k} &\rightarrow &r_{j}(\omega ,q,u)\equiv \frac{f_{j}(\omega ,q,u)\omega
^{2}-(qv_{j}u)^{2}}{\omega ^{2}-(qv_{j}u)^{2}}  \label{multiband r}
\end{eqnarray}

For \textit{n} bands, the effective pairing potential is assumed to be
represented by a symmetric \textit{n} \textsf{x}\ \textit{n} matrix $%
\boldsymbol{D}^{(\mathit{p})}$ with elements $D_{j,j^{\prime }}^{(p)}$ when $%
k$ is in band $j$ and $k^{\prime }$ is in band $j^{\prime }$. Let the
symmetric \textit{n} \textsf{x}\ \textit{n} matrix $\boldsymbol{D}^{(\mathit{%
v})}$ denote the effective potential in the p/h channel responsible for
vertex corrections and allow for the possibility that $\boldsymbol{D}^{(%
\mathit{p})}\neq \boldsymbol{D}^{(\mathit{v})}.$ With these assumptions, the
integrals in Eqs.(\ref{Vert eq gamma 3}-\ref{Bubble}) simplify into sums
over band index $j$\ of integrals over the Fermi surface of that band, where
the only variable in the integrand is $u$. The integral equations (\ref{Vert
eq gamma 3}) and (\ref{Vert eq gamma2}) turn into discrete equations for the
dressed vertices%
\begin{equation}
\text{ }\Gamma _{k}^{\left( 2,3\right) }\rightarrow \Gamma _{j}^{\left(
2,3\right) }(\omega ,q)  \label{renamed gammas}
\end{equation}

Upon introduction of the density of states $\rho _{j}$ for band $j$, the
integrals appearing on the right side of Eqs.(\ref{Vert eq gamma 3}-\ref%
{Bubble}) reduce to $\rho _{j}$ multiplied by an average over $u$. Let $%
F_{j}(\omega ,q),$ $P_{j}(\omega ,q),$ and $Q_{j}(\omega ,q)$ denote these
averages of $f_{j}(\omega ,q,u)$, $p_{j}(\omega ,q,u)$, and $r_{j}(\omega
,q,u)$, respectively. Make \textit{n}\ vectors $\boldsymbol{F}(\omega ,q)$,\ 
$\boldsymbol{P}(\omega ,q)$,\ and $\boldsymbol{Q}(\omega ,q)$\ out of them,
as well as \textit{n}\ vectors out of $\rho _{j},$ $\Delta _{j},$ $L_{j,}$
bare vertices $\gamma _{j}$ and dressed vertices $\Gamma _{j}^{(2,3)}(\omega
,q)$. For simplicity, we will often drop the explicit dependence on $(\omega
,q)$. Let $\boldsymbol{I}$ denote the \textit{n}\ dimensional unit matrix.
The dot \ $\boldsymbol{\cdot }$\ denotes the scalar product
of two vectors or matrix multiplication.

The existence of Leggett's mode requires the existence of the inverse\cite%
{comment}%
\begin{equation}
\boldsymbol{W}\equiv (\boldsymbol{D}^{(\mathit{p})})^{-1}.  \label{defn W}
\end{equation}

The following matrix equations were obtained. The explanation immediately
follows.\bigskip 
\begin{equation}
\boldsymbol{\Gamma }^{(3)}=\boldsymbol{\gamma }+\boldsymbol{D}^{(v)}\cdot 
\frac{i\omega }{2\boldsymbol{\Delta }}\left[ \boldsymbol{F\Gamma }^{(2)}-%
\frac{2i\boldsymbol{\Delta }}{\omega }\boldsymbol{Q}\mathbf{\Gamma }^{(3)}%
\right] \boldsymbol{\rho }  \label{matrix vert eq, gamma3}
\end{equation}%
\begin{equation}
\boldsymbol{W\cdot \Gamma }^{(2)}=-\boldsymbol{L\rho \Gamma }-\frac{\omega
^{2}}{4\boldsymbol{\Delta }^{2}}\left[ \boldsymbol{P}\mathbf{\Gamma }^{(2)}-%
\frac{2i\boldsymbol{\Delta }}{\omega }\boldsymbol{F}\mathbf{\Gamma }^{(3)}%
\right] \boldsymbol{\rho }  \label{matrix vert eq gamma2}
\end{equation}%
\begin{equation}
\boldsymbol{W\cdot \Delta }=-\boldsymbol{L\rho \Delta }
\label{matrix gap eq}
\end{equation}%
\begin{equation}
B_{\gamma ^{\ast },\Lambda \gamma }(\omega ,q)=-2\boldsymbol{\gamma }%
^{\prime }\cdot \frac{i\omega }{2\boldsymbol{\Delta }}\left[ \boldsymbol{%
F\Gamma }^{(2)}-\frac{2i\boldsymbol{\Delta }}{\omega }\boldsymbol{Q}\mathbf{%
\Gamma }^{(3)}\right] \boldsymbol{\rho }  \label{matrix bubble}
\end{equation}

Equations(\ref{matrix vert eq, gamma3}) and (\ref{matrix bubble}) are matrix
versions of Eqs.(\ref{Vert eq gamma 3}) and (\ref{Bubble}). Equations(\ref%
{matrix vert eq gamma2}) and (\ref{matrix gap eq}) were obtained by left
multiplying the matrix equations we obtain from Eqs.(\ref{Vert eq gamma2})
and (\ref{gap eqn}) by $\boldsymbol{W\cdot }$. Above, when we write, e.g., $%
\boldsymbol{L\rho \Gamma }^{(2)},$ we refer to the \textit{n}-vector with
components $L_{j}\rho _{j}\Gamma _{j}^{(2)}$.

Using the result $\boldsymbol{L\rho =}-\boldsymbol{\Delta }^{-1}\boldsymbol{%
W\cdot \Delta }$ from Eq. (\ref{matrix gap eq}) to eliminate $\boldsymbol{%
L\rho }$ from the first term on the right side of Eq.(\ref{matrix vert eq
gamma2}), we solve for $\boldsymbol{\Gamma }^{(2)}$ in terms of $\boldsymbol{%
\Gamma }^{(3)}$. After inserting the result in Eq.(\ref{matrix vert eq,
gamma3}) and solving for $\boldsymbol{\Gamma }^{(3)}(\omega ,q)$ in terms of 
$\boldsymbol{\gamma }$, we obtain from Eq.(\ref{matrix bubble}): 
\begin{equation}
B_{\gamma ^{\prime },\Gamma \gamma }(\omega ,q)=\boldsymbol{\gamma }^{\prime
}\cdot \boldsymbol{R}^{(p,v)}(\omega ,q)\cdot \boldsymbol{\gamma }\text{ .}
\label{resp matrix, pv}
\end{equation}%
Here $\boldsymbol{R}^{(p,v)}(\omega ,q)$ is the response-function matrix in
the presence of vertex corrections due to pairing $(p)$ and the vertex
correction $(v)$ due to the residual interaction in the particle/hole
channel,%
\begin{equation}
\boldsymbol{R}^{(p,v)}(\omega ,q)=\frac{\boldsymbol{I}}{\left[ \boldsymbol{R}%
^{(p)}(\omega ,q)\right] ^{-1}+\frac{1}{2}\boldsymbol{D}^{(v)}},
\label{eq for R(pv)}
\end{equation}%
\newline
where $\boldsymbol{R}^{(p)}(\omega ,q)$,\ is the response function matrix
with a correction only due to pairing: 
\begin{equation}
\boldsymbol{R}^{(p)}=-2\left[ \left\Vert \boldsymbol{Q\rho }\right\Vert
+\left\Vert \boldsymbol{F\rho }\right\Vert \cdot \frac{\boldsymbol{I}}{\frac{%
4}{\omega ^{2}}\boldsymbol{U}-\left\Vert \boldsymbol{P\rho }\right\Vert }%
\cdot \left\Vert \boldsymbol{F\rho }\right\Vert \right] .  \label{eq for Rp}
\end{equation}%
\newline
\newline
Above, dependence on $(\omega ,q)$\ remains understood, and $\left\Vert 
\boldsymbol{Q\rho }\right\Vert $, $\left\Vert \boldsymbol{F\rho }\right\Vert 
$ and $\left\Vert \boldsymbol{P\rho }\right\Vert $ denote \textit{n} \textsf{%
x}\ \textit{n} \ diagonal matrices with components $Q_{j}(\omega ,q)\rho
_{j} $, etc., on the diagonal. The symmetric matrix 
\begin{equation}
\boldsymbol{U\equiv }-\boldsymbol{\Delta }\left( \boldsymbol{W\Delta }%
-\left( \boldsymbol{W}\cdot \boldsymbol{\Delta }\right) \boldsymbol{I}%
\right) ,  \label{def of U}
\end{equation}%
\newline
or, in component form

\begin{equation}
U_{ij}=-\Delta _{i}\left( W_{ij}\Delta _{j}-\sum_{k=1}^{n}W_{ik}\Delta
_{k}\delta _{ij}\right)  \label{comp version for U}
\end{equation}%
imprints on the matrices $\boldsymbol{R}^{(p)}(\omega ,q)$ and $\boldsymbol{R%
}^{(p,v)}(\omega ,q)$ the effect of the vertex correction due to pairing. $%
\boldsymbol{U}$ has the property that the elements in a given row or column
sum to zero. Recalling $\boldsymbol{1}$, the \textit{n-}vector consisting of 
\textit{1's,} this property implies $\boldsymbol{U}\cdot \boldsymbol{1}=%
\boldsymbol{1}\cdot \boldsymbol{U}=0$.

Our fourth correction for final state interactions will result by summing
bubble diagrams consisting of the response function matrix $\boldsymbol{R}%
^{(p,v)}(\omega ,q)$ connected by a symmetric interaction matrix that we
will denote by $\boldsymbol{D}^{(b)}$.\cite{bubble} For clarity, we consider
the possibility that $\boldsymbol{D}^{(b)}\neq \boldsymbol{D}^{(v)}\neq 
\boldsymbol{D}^{(p)}$. The series $\boldsymbol{R}^{(p,v)}(\omega ,q)+%
\boldsymbol{R}^{(p,v)}(\omega ,q)\cdot \boldsymbol{D}^{(b)}\cdot \boldsymbol{%
R}^{(p,v)}(\omega ,q)+...$\ may be summed. The result, together with Eq.(\ref%
{eq for R(pv)}), gives a new response function matrix $\boldsymbol{R}%
^{(p,v,b)}(\omega ,q)$ obeying: 
\begin{equation}
\boldsymbol{R}^{(p,v,b)}=\frac{\boldsymbol{I}}{\left[ \boldsymbol{R}^{(p)}%
\right] ^{-1}+\frac{1}{2}\boldsymbol{D}^{(v)}-\boldsymbol{D}^{(b)}}.
\label{eq for Rpvb}
\end{equation}

The screened response function $\chi _{\gamma }(\omega ,q)\equiv B_{\gamma
^{\ast },\Lambda \gamma }(\omega ,q)$ then becomes 
\begin{equation}
\chi _{\gamma }(\omega ,q)=\boldsymbol{\gamma }^{\ast }\cdot \boldsymbol{%
\boldsymbol{R}\cdot \gamma }-\frac{\left( \boldsymbol{\gamma }^{\ast }%
\boldsymbol{\cdot \boldsymbol{R}\cdot 1}\right) \left( \boldsymbol{1}\cdot 
\boldsymbol{\boldsymbol{R}}\cdot \boldsymbol{\gamma }\right) }{\boldsymbol{1}%
\cdot \boldsymbol{\boldsymbol{R}}\cdot \boldsymbol{1}},
\label{screened bubble with R(p,v,b)}
\end{equation}

with $\boldsymbol{R}=\boldsymbol{R}^{(p,v,b)}(\omega ,q)$.

As discussed in the Appendix, the screening correction, the second term,
vanishes as $q\rightarrow 0$.

\section{The 2 band case with zero wave vector}

The remaining discussion concerns the case of \textit{n}=2. Using the above
assumption that $\boldsymbol{D}^{(p)}$ has an inverse (and thus the
determinant $D_{11}^{(p)}D_{12}^{(p)}-(D_{12}^{(p)})^{2}$ is nonzero), we
write the $q=0$ result for 2 bands as follows. (It can also be written in
the form of the expression given in Eqs. 3, 5 and 6 of Blumberg \textit{et
al.,\cite{Blumbergelectrons},} but generalized to include the bubble
diagrams.)%
\begin{align}
\chi _{\gamma }(\omega ,q& =0)=\frac{-2|\gamma _{2}-\gamma _{1}|^{2}}{\rho
_{1}^{-1}f_{1}^{-1}+\rho _{2}^{-1}f_{2}^{-1}-h\omega ^{2}+d^{(v)}-d^{(b)}},
\label{q=0 susc} \\
& f_{1,2\equiv }f(\frac{\omega }{2\Delta _{1,2}}),  \label{f12} \\
f(x)& \equiv \frac{\arcsin (x)}{x\sqrt{1-x^{2}}},  \label{f(omega)} \\
h& \equiv -\frac{D_{11}^{(p)}D_{12}^{(p)}-(D_{12}^{(p)})^{2}}{%
4D_{12}^{(p)}\Delta _{1}\Delta _{2}},  \label{h} \\
d^{(v)}& \equiv -(D_{11}^{(v)}+D_{22}^{(v)}-2D_{12}^{(v)}),  \label{d(v)} \\
d^{(p)}& \equiv -2(D_{11}^{(b)}+D_{22}^{(b)}-2D_{12}^{(b)}).  \label{d(p)}
\end{align}

\textit{The screening correction gives zero, as it must for} $q=0$. See the 
\textbf{Appendix}.

It is a convenient, but incorrect, assumption that final state interactions
can be neglected. Almost all authors do not make any corrections: pairing,
vertex, or bubble, represented above by nonzero $D_{ij}^{(p)}$, $%
D_{ij}^{(v)} $, and $D_{ij}^{(b)}$.\cite{exception} When these quantities
are set equal to zero, we obtain%
\begin{equation}
\chi _{\gamma }^{0}(\omega ,q=0)=\frac{-2|\gamma _{2}-\gamma _{1}|^{2}}{\rho
_{1}^{-1}f_{1}^{-1}+\rho _{2}^{-1}f_{2}^{-1}}.  \label{bare susc}
\end{equation}%
This is identical to the result that would be obtained by using the
expression \textit{with} screening, Eq.(\ref{screened bubble with R(p,v,b)}%
), but with $\boldsymbol{\boldsymbol{R}^{(p,v,b)}}(\omega ,0)$ replaced by
the bare response function matrix $\boldsymbol{\boldsymbol{R}^{0}}(\omega
,0)=-2\left\Vert \boldsymbol{F}(\omega ,0)\boldsymbol{\rho }\right\Vert $.
The difference between $\chi _{\gamma }(\omega ,q=0)$\ and $\chi _{\gamma
}^{0}(\omega ,q=0)$\ is both quantitative and qualitative.

\section{Application to Fe-pnictides}

Whereas the $q=0$ assumption is not realistic for $MgB_{2}$, it is for the
Fe-pnictides, magnetic superconductors with a layered structure that makes
them nearly two-dimensional, with very small components of the Fermi
velocities perpendicular to the layers. For them, we apply Eqs.(\ref{q=0
susc}) through (\ref{d(p)}) to the two-band model \ with extended s-wave
symmetry considered by Chubukov \textit{et al.} \cite{Andrey}. Those authors
assume that $\Delta _{2}=-\Delta _{1}=-\Delta $ and $\rho _{1}=\rho
_{2}=\rho .$ In their notation, our matrices describing pairing, vertex, and
bubble corrections would read:%
\begin{eqnarray}
\boldsymbol{D}^{(p)} &=&\rho ^{-1}\left[ 
\begin{array}{cc}
u_{4} & u_{3} \\ 
u_{3} & u_{4}%
\end{array}%
\right] ;\text{ }\boldsymbol{D}^{(v)}=\rho ^{-1}\left[ 
\begin{array}{cc}
u_{4} & u_{2} \\ 
u_{2} & u_{4}%
\end{array}%
\right] ;  \notag \\
\text{ }\boldsymbol{D}^{(b)} &=&\rho ^{-1}\left[ 
\begin{array}{cc}
u_{4} & u_{1} \\ 
u_{1} & u_{4}%
\end{array}%
\right] .  \label{D matrices u's}
\end{eqnarray}

We then find%
\begin{align}
\chi _{\gamma }(\omega ,q& =0)=\frac{-|\gamma _{2}-\gamma _{1}|^{2}\rho
f\left( \frac{\omega }{2\Delta }\right) }{1-\left( u_{eff}+\frac{\omega
^{2}\left( u_{4}^{2}-u_{3}^{2}\right) }{8\Delta ^{2}u_{_{3}}}\right) f\left( 
\frac{\omega }{2\Delta }\right) },  \label{Andrey's result per me} \\
u_{eff}& \equiv 2u_{1}-u_{2}-u_{4}.  \notag
\end{align}

Using renormalization group arguments, Chubukov \textit{et al.} asserted
that $u_{eff}$ is dominated by $u_{1}$ and is positive. This means that
among all final state interactions, the dominant one is repulsive, and it
couples p/h pairs in one band to those in the other band. They assigned the
value 0.4 to $u_{eff}$. If $\frac{\omega ^{2}\left(
u_{4}^{2}-u_{3}^{2}\right) }{8\Delta ^{2}u_{_{3}}}$ is neglected with
respect to $u_{eff},$ our result becomes similar to theirs:%
\begin{equation}
\chi _{\gamma }(\omega ,q=0)=-\frac{|\gamma _{2}-\gamma _{1}|^{2}\rho
f\left( \frac{\omega }{2\Delta }\right) }{1-u_{eff}f\left( \frac{\omega }{%
2\Delta }\right) }.  \label{A r p m with further assump.}
\end{equation}%
\newline
When plotted, $\textrm{Im}\chi _{\gamma }(\omega +i0^{+},0)$\ for $u_{eff}=0.4$
is similar to the plot shown in their Fig. 1b. There is a difference. Based
on an argument involving an energy-dependent cut-off of the effective
interaction, they used $f\left( \frac{\omega }{2\Delta }\right) -1$ in place
of our $f\left( \frac{\omega }{2\Delta }\right) $. This replacement would
give incorrect results if applied to the Ideal Limit to be discussed next.

\section{Two-Band Case with finite wave vector in the Ideal Limit}

Following Leggett,\cite{Tony} we consider the case where the material
parameters and regions of interest are such that $\omega $ and $qv_{j}$ are
both assumed to be non-zero but small compared with $\Delta _{j}$. This
leads to the approximations%
\begin{equation}
F_{j}\cong Q_{j}\cong 1\text{ ; }P_{j}\cong 1-\left( \frac{qc_{j}}{\omega }%
\right) ^{2};  \label{F, Q, P in ideal limit}
\end{equation}%
where%
\begin{equation}
c_{j}^{2}\equiv \mathit{v}_{j}^{2}\left\langle u_{k_{j}}^{2}\right\rangle
_{FS}=\frac{\mathit{v}_{j}^{2}}{3}.  \label{def of cj}
\end{equation}%
The last equality results from the assumption that the Fermi surface is
spherical. With these assumptions, Eqs.($\ref{eq for Rpvb}$ and \ref{eq for
Rp}) lead to the result%
\begin{equation}
\boldsymbol{R}^{(p,v,b)}(\omega ,q)=\frac{-2}{\left( \left\Vert \boldsymbol{%
\rho }^{-1}\right\Vert -\frac{\boldsymbol{I}\omega ^{2}}{4\boldsymbol{U}%
+q^{2}\left\Vert \boldsymbol{c}^{2}\boldsymbol{\rho }\right\Vert }\right) -%
\boldsymbol{D}^{(v)}+2\boldsymbol{D}^{(b)}}.  \label{Inv R2 in ideal limit}
\end{equation}%
\ Recall that $\left\Vert \boldsymbol{\rho }^{-1}\right\Vert $ and $%
\left\Vert \boldsymbol{c}^{2}\boldsymbol{\rho }\right\Vert $ denote diagonal
matrices. This gives the following results from Eq.(\ref{screened bubble
with R(p,v,b)}): 
\begin{subequations}
\begin{equation}
\chi _{\gamma }(\omega ,q)=B_{\gamma ^{\ast },\Lambda \gamma }(\omega ,q)=%
\frac{8|\gamma _{2}-\gamma _{1}|^{2}\left( \omega
_{o}^{2}+v_{o}^{2}q^{2}\right) /V_{o}}{\omega ^{2}-\omega _{o}^{2}-\
q^{2}v_{o}^{2}},  \label{Bubble result, ideal limit}
\end{equation}%
\begin{equation}
\omega _{o}^{2}\equiv -\frac{V_{o}8D_{12}^{(p)}\Delta _{1}\Delta _{2}}{%
D_{11}^{(p)}D_{22}^{(p)}-(D_{12}^{(p)})^{2}},  \label{Def omega_o}
\end{equation}%
\begin{equation}
v_{o}^{2}\equiv \frac{V_{o}\rho _{1}\rho _{2}c_{1}^{2}c_{2}^{2}}{\rho
_{1}c_{1}^{2}+\rho _{2}c_{2}^{2}},  \label{Dev v_o}
\end{equation}%
\begin{equation}
V_{o}\equiv \frac{1}{\rho _{1}}+\frac{1}{\rho _{2}}+d^{(v)}-d^{(b)}.
\label{Def V_o}
\end{equation}

[Recall Eqs.(\ref{d(v)}) and (\ref{d(p)})]. We derive from the denominator
of Eq.(\ref{Bubble result, ideal limit}) the existence of a collective mode
at an energy $\omega =\omega _{L}(q)$ obeying 
\end{subequations}
\begin{equation}
\omega _{L}^{2}(q)\equiv \omega _{o}^{2}+v_{o}^{2}q^{2}
\label{LM dispersion}
\end{equation}%
with an energy (\lq mass\rq) $\omega _{o}$ and a dispersion velocity v$_{o}.$
These are generalizations of the results given by Leggett \cite{Tony} in his
Eqs.(3.39). If our parameter $V_{o}$ were given by $V_{o}\equiv \rho
_{1}^{-1}+\rho _{2}^{-1}$, and taking into account the fact that our density
of states $\rho _{i}$ is for one spin, whereas his is for both spins, we
would obtain his result. Thus, setting $\boldsymbol{D}^{(v)}=\boldsymbol{D}%
^{(b)}=0,$ i.e., the absence of a vertex correction in the particle/hole
channel and of a bubble correction, gives his results for the mode frequency
and its dispersion with $q$.

Use of Eqs.(\ref{q-integrated int}) and (\ref{Bubble result, ideal limit})
gives us an analytical result for the Raman intensity in this ideal case. If 
$\omega \leq \omega _{o}$, $I_{ideal}(\omega )=0$. If $\omega >\omega _{o}$, 
\begin{subequations}
\begin{align}
I_{ideal}(\omega )& =\frac{S_{fqi}16q^{\prime \prime }\omega ^{2}\Phi
(q^{\prime },q^{\prime \prime },\omega )}{(\rho _{1}+\rho _{2})V_{o}v_{o}%
\sqrt{\delta \omega ^{2}}},  \label{I(Omega)} \\
\Phi (q^{\prime },q^{\prime \prime },\omega )& \equiv \frac{q^{\prime
2}+q^{\prime \prime 2}+v_{o}^{-2}\delta \omega ^{2}}{\left[ q^{\prime
2}+q^{\prime \prime 2}+v_{o}^{-2}\delta \omega ^{2}\right]
^{2}-4v_{o}^{-2}\delta \omega ^{2}q^{\prime 2}},  \label{delta(etc)} \\
\delta \omega ^{2}& \equiv \omega ^{2}-\omega _{o}^{2}\text{, }S_{fqi}\equiv 
\frac{|\gamma _{2}-\gamma _{1}|^{2}(\rho _{1}+\rho _{2})}{4q^{\prime \prime }%
}\text{.}  \label{SF_q_int}
\end{align}%
\linebreak $S_{fqi}$ is introduced as a scale factor for this situation
where an integral has been performed over $q$. The quantity $%
I_{ideal}(\omega )/S_{fqi}$ is dimensionless.

\section{Application to Magnesium Diboride}

For $MgB_{2}$ the superconducting pairing comes primarily from optical
phonons and is attractive.\cite{phonons}\ The intra $\sigma $ band effecting
pairing potential $D_{11}^{(p)}$ is due to modes that become the doubly
degenerate optical phonon of $E_{g}$ symmetry at $q\approx 0$ wherein the
two $B$ atoms in the unit cell move perpendicular to the $c$ axis. The intra 
$\pi $ band pairing potential $D_{22}^{(p)}$\ and inter $\pi -\sigma $ band
pairing potential $D_{12}^{(p)}$\ are due to the mode that becomes the
optical phonon of $B_{1g}$ symmetry at $q\approx 0$ wherein the $B$ atoms
move parallel to the $c$ axis. Neither of these phonons for the relatively
small (with respect to the inverse lattice constant) physical values of $q$
are capable of making a contribution to the bubble correction, because the
lines connecting the bubbles are, in fact proportional to phonon propagators
and they must have the same symmetry, $A_{1g}$, as that of the bare Raman
vertex. Thus $\boldsymbol{D}^{(b)}$ can only be due to the much weaker
effects of $A_{1g}$ symmetry acoustical phonons. We thus assume $\boldsymbol{%
D}^{(b)}=0$. On the other hand, for pairing and for the vertex correction,
no such symmetry arguments apply, and we simply assume that $\boldsymbol{D}%
^{(v)}=\boldsymbol{D}^{(p)}=\boldsymbol{D}.$

\begin{table}[pbt] \centering

\begin{tabular}{||l|l|l|l|l|l|l||}
\hline\hline
Name & $D_{11}$ & $D_{22}$ & $D_{12}$ & $\rho _{1}$ & $\rho _{2}$ &  \\ 
\hline
Value & -0.47 & -0.1 & -0.08 & 2.04 & 2.78 &  \\ \hline
Units & Ry-c & Ry-c & Ry-c & /Ry-spin-c & /Ry-spin-c &  \\ \hline\hline
Name & $\Delta _{1}$ & $\Delta _{2}$ & $v_{1}$ & $v_{2}$ & $q^{\prime }$ & $%
q^{\prime \prime }$ \\ \hline
Value & 6.75 & 2.3 & 0.38 & 4.7 & 1.73 & 4.18 \\ \hline
Units & meV & meV & 10$^{5}$m/s & 10$^{5}$m/s & 10$^{7}$m$^{-1}$ & 10$^{7}$m$%
^{-1}$ \\ \hline\hline
\end{tabular}%
\caption{Values of input parameters used in the calculation for MgB$_{2}$. Here \lq Ry\rq\, denotes the Rydberg, and \lq c\rq\, means \lq unit cell.\rq\,
To convert the units of the product of $q^{\prime \prime}$ with
$v_{1}$  into meV, multiply by $\hbar=0.658\cdot10^{-12}$ meV s.} 

\end{table}%

The material parameters used by Blumberg \textit{et al.}\cite%
{Blumbergelectrons}\textit{\ }are those shown inTable I. The values of $%
q^{\prime \prime }$ and $q^{\prime }$ are estimates based on optical data. 
\cite{optics} The other parameters are from Liu et al. \cite{Liu}. Using
these parameters we calculate $\omega _{o}=12.42$ meV and $v_{o}=0.23$ $%
\cdot $10$^{5}$m/s. This value of $\omega _{o}$ is much too large for the
conditions of the ideal limit to be satisfied. Note that roughly $\omega
_{o}\propto \sqrt{-D_{12}}$ and that $v_{o}$ depends very weakly on $%
-D_{12}\ $as long as $-2D_{12}\ll \rho _{1}^{-1}+\rho
_{2}^{-1}-D_{11}-D_{22} $, which is the case here. For example, if $-D_{12}\ 
$were to equal $0.001$ Ry-cell, we would have $\omega _{o}=1.37$ meV, less
than $2\Delta _{2}$, and the ideal limit would amount to a good
approximation.

\section{The Real Case}

We now turn to the more realistic case, where $F_{j},$ $Q_{j},$ and $P_{j}$
have complicated dependences on $\omega $ and $q$. They depend on the
properties of $\ f(\beta )\equiv \frac{\arcsin \beta }{\beta \sqrt{1-\beta
^{2}}}$ as a function of $\beta =\beta _{j}(q,u,\omega )\equiv \sqrt{\frac{%
\omega ^{2}-(qv_{j}u\ )^{2}}{4\Delta _{j}^{2}}}$. Keeping in mind that $%
\omega $ has an infinitesimally small positive imaginary part, we may write
for various regions: 
\end{subequations}
\begin{subequations}
\begin{align}
f(\beta )& =-\frac{\ln \left( \sqrt{1+|\beta |^{2}}-|\beta |\right) }{|\beta
|\sqrt{1+|\beta |^{2}}},\text{ when }\beta ^{2}<0,
\label{f(beta) in region 0} \\
f(\beta )& =\frac{\arcsin (\beta )}{\beta \sqrt{1-\beta ^{2}}}  \notag \\
& =\frac{\frac{\pi }{2}-\arcsin (\sqrt{1-\beta ^{2}})}{\beta \sqrt{1-\beta
^{2}}},\text{ when }0\leq \beta <1,  \label{f(beta) in regions 1,2} \\
f(\beta )& =\frac{\ln (\beta -\sqrt{\beta ^{2}-1)})}{\beta \sqrt{\beta ^{2}-1%
}}+\frac{\pi i/2}{\beta \sqrt{\beta ^{2}-1}},\text{ when }\beta >1.
\label{f(beta) in region 3}
\end{align}

These expressions are to be inserted into Eqs.(\ref{multiband p}) and (\ref%
{multiband r}). We replace $\beta $ with $\left( 2\Delta _{j}\right) ^{-1}%
\sqrt{\omega ^{2}-(qv_{j}u\ )^{2}}$ and then average over $u$ to obtain $%
F_{j}(q,\omega ),$ $Q_{j}(q,\omega ),$ and $P_{j}(q,\omega ).$ The $u$%
-integrals yield elliptic integrals for the imaginary parts, but we know of
no analytic forms for the real parts. We chose not to rely on numerical
integration to obtain the real parts of $F,P,$ and $Q.$ Instead we
approximate the real parts of $f,p,$ and $r$ with functions of $\beta $ that
can be averaged over $u$. We used these the following approximate functions
for the real part of $f(\beta )$. For $\beta ^{2}<0$, $0\leq \beta \leq 
\frac{1}{\sqrt{2}}$, $\frac{1}{\sqrt{2}}\leq \beta <1$, and $\beta >1$,
respectively, we used: 
\end{subequations}
\begin{subequations}
\begin{align}
f^{\left( 0\right) }\left( \beta \right) & \equiv \frac{7}{3\left( 1-\beta
^{2}\right) }-\frac{3}{\left( 1-\beta ^{2}\right) ^{2}}+\frac{5}{3\left(
1-\beta ^{2}\right) ^{4}},  \label{def f(0)} \\
f^{\left( 1\right) }\left( \beta \right) & \equiv \frac{1+\frac{1}{6}\beta
^{2}+\beta ^{4}\left( -\frac{13}{3}+\sqrt{2}\pi \right) }{\sqrt{1-\beta ^{2}}%
},  \label{def f(1)} \\
f^{\left( 2\right) }\left( \beta \right) & \equiv \frac{\frac{\pi }{2}}{%
\beta \sqrt{1-\beta ^{2}}}-f^{\left( 1\right) }\left( \sqrt{1-\beta ^{2}}%
\right) ,  \label{def f(2)} \\
f^{\left( 3\right) }\left( \beta \right) & \equiv \frac{-\frac{5}{6}-\frac{1%
}{2}\ln (\beta ^{2})}{\beta ^{2}}-\frac{1}{6\beta ^{4}}.  \label{def f(3)}
\end{align}

These have the properties that $f^{\left( 0\right) }\left( \beta \right) $
and $f^{\left( 1\right) }\left( \beta \right) $\ tend to $1$ as $\beta ^{2}$
tends to $0$ from below and above, respectively; $f^{\left( 2\right) }\left(
\beta \right) $ has the same limiting, singular behavior as $f(\beta )$,
when $\beta $ tends to $1$ from below; and $f^{\left( 3\right) }\left( \beta
\right) $ tends to the correct limit, $-1,$ when $\beta $ tends to $1$ from
above. Moreover, the approximations $f^{\left( 1\right) }\left( \beta
\right) $ and $f^{\left( 2\right) }\left( \beta \right) $ agree when $\beta
=1/\sqrt{2}$. These approximations for $f(\beta )$ were inserted in Eqs.(\ref%
{multiband p}) and (\ref{multiband r}) to obtain approximate, integrable,
versions of the real parts of $p(\beta )$ and $r(\beta )$. The averages over 
$u$ were than calculated analytically with the help of \lq Mathematica.\rq\, The
resulting functions $F_{j},$ $P_{j},$ and $Q_{j}$ of $q$ and $\omega $\ were
used in Eqs.(\ref{eq for Rp}), (\ref{eq for Rpvb}), and (\ref{screened
bubble with R(p,v,b)}) to calculate analytically $\textrm{Im}[B_{\Lambda
(\gamma ),\gamma ^{\ast }}(q,\omega +i0^{+})]=\textrm{Im}[\chi _{\gamma
}(q,\omega +i0^{+})]$ that was inserted into Eq.(\ref{q-integrated int}) for
numerical integration over $q$ from $0$ to an upper limit of $10q^{\prime
\prime }$ that was sufficiently large to be accurate in the peak region of
the intensity plots.

\section{Numerical Results}

We start by comparing in Fig.(\ref{fig1}) the ideal case results for $%
-D_{12}=0.001,$ $0.003,$ and $0.01$\ with the real case result for the
unrealistically small value of $-D_{12}=0.01.$

\begin{figure}[hbt]
\begin{center}
\includegraphics[width=0.5\textwidth]{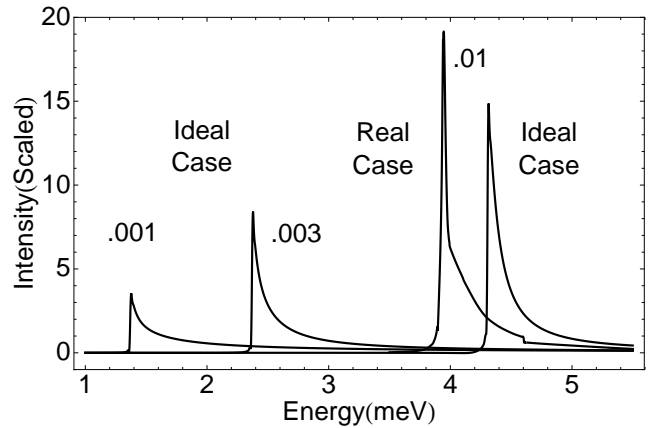}
\end{center}
\caption{Scaled Raman intensity for the
ideal case with inter-band pairing potential $-D_{12}=0.001,$ $0.003,$ and $%
0.01$. The singularity at each peak was rounded. Also shown is the real case
result for $-D_{12}=0.01$. Here, a small positive imaginary part was added
to the energy variable $\omega $.}
\label{fig1}
\end{figure}%

Note the similarity of the two $-D_{12}=0.01$\ spectra. Note also the
discontinuity at $\omega =2\Delta _{2}=4.6$ meV for the real case. For more
realistic values of $-D_{12}$ the effect of departure of the functions $%
F_{j},$ $P_{j},$ and $Q_{j}$ from their ideal case forms leads to dramatic
changes in the Raman spectra, as shown in Fig.(\ref{fig2}). Note that the
discontinuity at $\omega =2\Delta _{2}$ is now joined by one at $\omega
=2\Delta _{1}=13.5$ meV.

\begin{figure}[hbt]
\begin{center}
\includegraphics[width=0.5\textwidth]{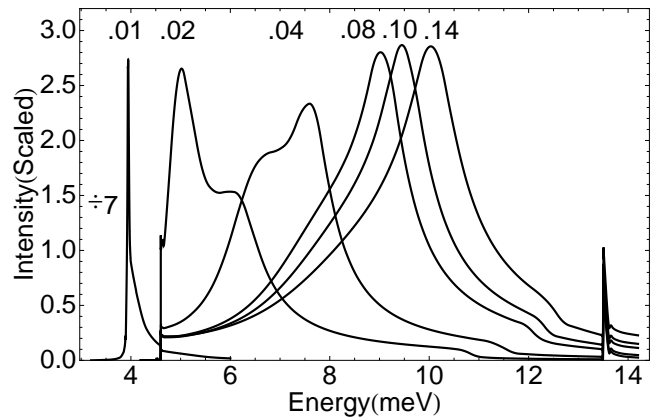}
\end{center}
\caption{Scaled Raman intensity for the
real case for $-D_{12}=0.01,$ $0.02,$ $0.04,$ $0.08,$ $0.10,$ and $0.14$.
For the 0.01 plot a small positive imaginary part was added to the energy
variable $\omega $. \lq Wiggles\rq\, near 4.6 and 13.5 meV, the values of the 2
gaps, $2\Delta _{1}$and $2\Delta _{2}$, are artifacts of smoothing the
numerical integration carried out at a discreet number of points.}%
\label{fig2}
\end{figure}%

Table I gives a calculated value for $-D_{12}$\ of 0.08. The experimental
result of of Blumberg \textit{et al.} \cite{Blumbergelectrons} shows that
the peak occurs at 9.4 meV. Our calculation with $-D_{12}=0.08$ gives the
peak energy at 9.0 meV, whereas our calculation with $-D_{12}=0.10$ has a
peak energy very close to the experimental value. Taken together, the
results in Figs. (1-2) show how the collective mode, first predicted by
Leggett, but corrected for final state interactions, evolves into a
resonance mode peaked very close to the energy found by the Raman experiment
as the parameter $-D_{12}$\ increases from very small values to those close
to the LDA result.

We show a final comparison in Fig.(\ref{fig3}b), namely between the $%
-D_{12}=0.10$ result from Fig.(\ref{fig2}) and the result with the same
value of $-D_{12}$\ obtained by setting $q=0$. The effect of finite values
of $q$ on shifting up the energy of the peak is obvious. This is a kinematic
effect that crudely results from the need for higher values of $\omega $ to
compensate for nonzero values of $\left( q\cdot v_{k}\right) ^{2}$ in Eqs.(%
\ref{Def beta, f}), (\ref{multiband p}) and (\ref{multiband r}).

These effects play out in an interesting way in our case where the 2 Fermi
velocities differ by a factor of 12.\ The peak of -$\textrm{Im}[\chi _{\gamma
}(q,\omega )]$ moves from 8 meV at $q=0$ to \textit{lower} energy as $q$
increases from $0$ to about $2\cdot 10^{7}$ m$^{-1}$. Then the peak moves to
higher energy. The behavior of the integrand in Eq.(\ref{q-integrated int})
can be seen in the density plot in Fig.(\ref{fig3}a).

\begin{figure}[hbt]
\begin{center}
\includegraphics[width=0.48\textwidth]{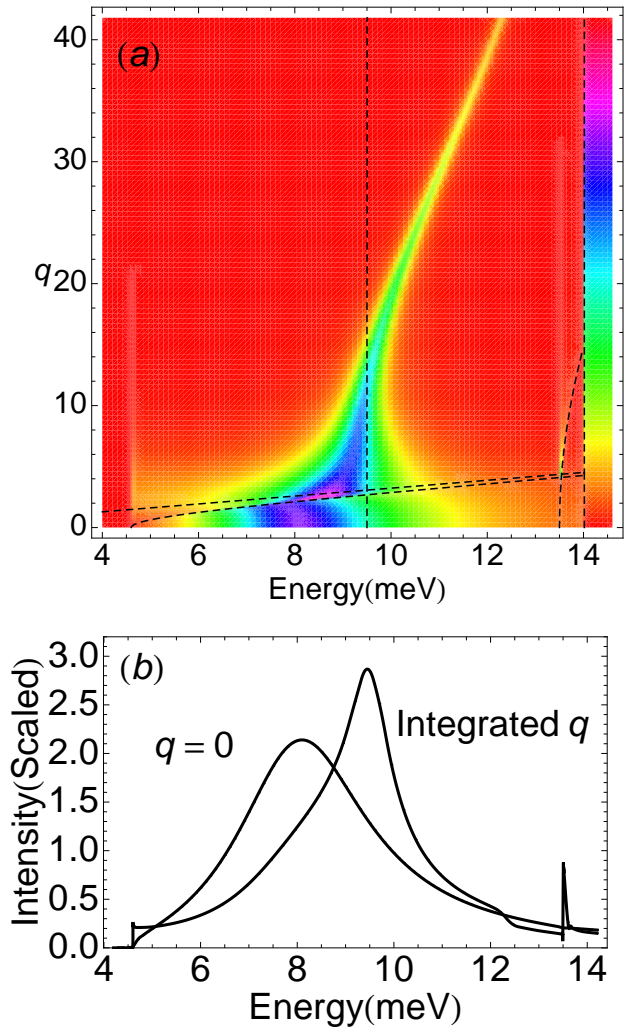}
\end{center}
\caption{(Color online) Results for the
interband pairing potential $-D_{12}=0.10$. (a) Density plot of the
integrand in Eq.(11) with values of $q$ (in units of $10^{7}$ $m^{-1}$)
extending to $10q^{\prime \prime }$. Intensity scale is shown by the color
bar on the right. The dashed curve starting at 4.6 meV obeys \ $\hslash
v_{\pi }k=\sqrt{\omega ^{2}-4\Delta _{\pi }^{2}}$. The dashed slanted
straight line obeys $\hslash v_{\pi }k=\omega ,$ and the dashed curve
starting at 13.5 meV obeys $\hslash v_{\sigma }k=\sqrt{\omega ^{2}-4\Delta
_{\sigma }^{2}}$. (b) The \lq Integrated q\rq\, plot is the same as the $%
-D_{12}=0.10$ plot in Fig.(2). The vertical dashed line at 9.5 meV in (a)
denotes the value of Energy where the \lq Integrated q\rq\, plot has its maximum.
The \lq $q=0$\rq\, plot is equivalent to a $q=0$ cut along the lower edge of the
density plot in (a). }
\label{fig3}
\end{figure}%

Note that the shoulder at 12.3 meV in the \lq Integrated q\rq\, curve of Fig (\ref%
{fig3}b) corresponds to the peak at the $q=10q^{\prime \prime }$ cut at the
top of the density plot of Fig.(\ref{fig3}a). If we were to extend the upper
limit of the integral over $q$ beyond $10q^{\prime \prime }$, this shoulder
would move up in energy, but there would be no effect in the shape of the
\lq Integrated q\rq\, plot in the energy range between 4.6 and 11.5 meV. The
dispersion of the peak in the density plot with wavevector $q$ beyond about
10$^{8}$ m$^{-1}$ is caused by the non-zero, but small, nature of the
velocity $v_{\sigma }$. If we had set $v_{\sigma }=0$, the peak in the
\lq Integrated q\rq\, plot would occur at 9 meV, and it would be higher and
narrower than in the plot shown in Fig.(\ref{fig3}b).

Recall that we had set $\mathbf{D}^{(v)}=\mathbf{D}^{(p)}\neq 0$ and $%
\mathbf{D}^{(b)}=0$\ to obtain these results. Had we set $\mathbf{D}^{(v)}=0$
and $\mathbf{D}^{(b)}=0$, the resulting Raman intensity calculation would
give a peak at a lower energy. In particular, the peak features shown in the
the density plot in Fig.(\ref{fig3}a) would occur at energies 1.5-2 meV
lower. Had we used $\mathbf{D}^{(v)}=\mathbf{D}^{(p)}=\mathbf{D}^{(b)}\neq 0$%
, they would have occured at still lower energies.

\section{Discussion}

To the author's knowledge this work amounts to the first attempt at a full
treatment of finite wave vector effects in the calculation of electronic
Raman scattering in a superconductor. Our "real case" calculation takes
these effects into account only through the use of the toy model assumption
of spherical Fermi surfaces together with approximations to the real part of 
$f(\beta )$ [Eqs.(\ref{def f(0)}-\ref{def f(3)})]. A more realistic
calculation would involve a much more complex model of the Fermi surfaces
and a great deal more massive numerical calculation. Nevertheless, the toy
model used here seems to have captured the essential features of Leggett's
mode as applied to the real multiband superconductor MgB$_{2}$: damping due
to decay into the $\pi $ band(s) and finite wavevector effects in both $\pi $
and $\sigma $ bands.

When applied to the normal state, our model does not account quantitatively
or qualitatively for the experimental results in $A_{1g}$ symmetry.\cite%
{Blumbergelectrons} The data show a rising continuum in both $A_{1g}$ and $%
E_{2g}$ symmetries that is cut-off below $2\Delta _{\pi }$ in the
superconducting state. An explanation will require a model for the bare
vertex $\gamma _{k}$\ that goes beyond the scope of the present work.

\begin{acknowledgments}
The author thanks Girsh Blumberg for sharing his Raman data on MgB$_{2\text{ 
}}$and for his insight that the experimental A$_{1g}$ peak at 9.4 meV was
the manifestation of Leggett's collective mode in a real multiband
superconductor. He also thanks Tony Leggett for discussions and a key
question. The answer convinced the author that, in principle, the bubble
correction needs to be considered in the context of electronic Raman
scattering.
\end{acknowledgments}

\appendix

\section{Flow, Counter-Flow, and the Screening Correction}

Here we derive an important result that has rather general applicability. It
concerns the interplay between the correction for pairing and the screening
correction when the wavevector $q$ is zero. In that case we have 
\end{subequations}
\begin{equation*}
r_{k}(\omega ,0)=p_{k}(\omega ,0)=f_{k}(\omega ,0)\equiv \frac{\arcsin (%
\frac{\omega }{2\Delta _{k}})}{\frac{\omega }{2\Delta _{k}}\sqrt{1-(\frac{%
\omega }{2\Delta _{k}})^{2}}}.
\end{equation*}%
All the factors in square brackets in Eqs.(\ref{Vert eq gamma 3}), (\ref%
{Vert eq gamma2}) and (\ref{Bubble}) become $f_{k^{\prime }}(\omega
,0)[\Gamma _{k^{\prime }}^{\left( 2\right) }(\omega ,0)-2i\Delta _{k^{\prime
}}\omega ^{-1}\Gamma _{k^{\prime }}^{\left( 3\right) }(\omega ,0)]$.

To examine the implications for screening, we need to know the behavior of $%
\Gamma _{k,1}^{\left( 2,3\right) }(\omega ,0)$, where the second subscript
denotes the value of $\gamma _{k}=1$. We find that setting%
\begin{equation*}
\Gamma _{k,1}^{\left( 2\right) }(\omega ,0)=2i\Delta _{k}\omega ^{-1}\text{
and }\Gamma _{k,1}^{\left( 3\right) }(\omega ,0)=1,
\end{equation*}%
thus making $[\Gamma _{k^{\prime },1}^{\left( 2\right) }(\omega ,0)-2i\Delta
_{k^{\prime }}\omega ^{-1}\Gamma _{k^{\prime },1}^{\left( 3\right) }(\omega
,0)]=0$, trivially satisfies both Eq.(\ref{Vert eq gamma 3}) and [with the
help of Eq.(\ref{gap eqn})] Eq.(\ref{Vert eq gamma2}). These results show
that the term in square brackets vanishes for each $k$ on the right side of
Eq.(\ref{Bubble}) for $B_{\gamma ^{\prime },\Gamma 1}(\omega ,0)$. Thus, $%
B_{\gamma ^{\prime },\Gamma 1}(\omega ,0)=0$ for any $\gamma _{k}^{\prime }$.

Of the two terms in square brackets in Eq.(\ref{Bubble}), the $\Gamma
_{k}^{\left( 3\right) }$ term can be thought of as \lq flow\rq\, of a $k$-dependent
longitudinal particle/hole current, and the $\Gamma _{k}^{\left( 2\right) }$
term as a $k$-dependent particle/particle \lq counter-flow\rq\, current produced in
response to the p/h current. For each value of $k$, p/p and p/h currents
tend to cancel, and, when the p/h current is uniform in $k$ space ($%
\boldsymbol{\gamma }_{k}\propto 1$), we have just shown that this
cancellation is exact for each $k$. At a deeper level the result $\Gamma
^{(2)}(1)=2i\Delta \omega ^{-1}$ is a result of gauge invariance and
particle conservation. For example, see the first of Nambu's Eqs. (7.6).\cite%
{Nambu}

For non-zero $\boldsymbol{q}$, Eqs. (\ref{Vert eq gamma 3}), (\ref{Vert eq
gamma2}) and (\ref{Bubble}) apply with $p_{k}(\omega ,\boldsymbol{q}%
)-f_{k}(\omega ,\boldsymbol{q})$ and $r_{k}(\omega ,\boldsymbol{q}%
)-f_{k}(\omega ,\boldsymbol{q})\neq 0$, and for small $q$, the leading terms
in an expansion are $\propto q^{2}$. This has the consequence that the
leading term in an expansion of $B_{\gamma ^{\prime },\Gamma 1}(\omega ,%
\boldsymbol{q})$ is $\propto q^{2}$. This also applies to $B_{\gamma ^{\ast
},\Gamma 1}(\omega ,\boldsymbol{q})$, $B_{1,\Gamma \gamma }(\omega ,%
\boldsymbol{q})$, and $B_{1,\Gamma 1}(\omega ,\boldsymbol{q})$\ in the
screening correction. [See Eqs.(\ref{screened bubble} and \ref{screening
corr}).] Thus, the leading term in an expansion of the screening correction
is $\propto q^{2}$.

\textbf{We have shown that the screening correction gives zero in the }$q=0$%
\textbf{\ limit}. We have sketched the proof here for the case of
non-retarded interactions restricted to the Fermi surface. A similar proof
involving vertex corrections, bubble corrections, and screening should be
possible for non-retarded interactions not restricted to the Fermi surface.
Because of the underlying connection to gauge invariance, this statement
about the screening correction should hold with retarded interactions as
well.

Thus, for $A_{1g}$ symmetry and $q=0$, it is not correct to use the bare
bubble, Eq.(\ref{Bubble}), $B_{\gamma ^{\prime },\gamma }(\omega ,0)=-2\int
dS_{k}\gamma _{k}^{\prime }f_{k}(\omega ,0)\gamma _{k}$, in the bare version
of the screening correction, Eq.(\ref{screening corr}), with $\Gamma
_{1}\rightarrow 1$ and $\Gamma _{\gamma }\rightarrow \gamma $. One must have
a solvable model for the pairing interaction [$D^{(p)}(k,k^{\prime })$] and
the vertex correction [$D^{(v)}(k,k^{\prime })$], and then solve equations
like Eqs.(\ref{Vert eq gamma2}) and (\ref{Vert eq gamma 3}) to obtain
expressions for $\Gamma _{k,\gamma }^{\left( 2\right) }(\omega ,0)$ and $%
\Gamma _{k,\gamma }^{\left( 3\right) }(\omega ,0)$\ to use in Eq.(\ref%
{Bubble}). One way to do this would involve a re-interpretation of our
solution for the multiband case. One can interpret matrix Eqs.(\ref{defn W}-%
\ref{matrix bubble}) as resulting from a fine-scale, or large \textit{n},
discrete version of Eqs.(\ref{Vert eq gamma 3}-\ref{Bubble}). The solutions
are given by Eqs (\ref{resp matrix, pv}-\ref{eq for Rpvb}). The response
function matrices $\boldsymbol{R}^{(p)}(\omega ,q)$, $\boldsymbol{R}%
^{(p,v)}(\omega ,q)$, and $\boldsymbol{R}^{(p,v,b)}(\omega ,q)$ are
symmetric. They can be shown to have the property that $\boldsymbol{1}\cdot 
\boldsymbol{R}(\omega ,q)=\boldsymbol{R}(\omega ,q)\cdot \boldsymbol{1}%
\propto q^{2}$ for small $q$.

Another solvable model results from using the product ansatz $D_{jj^{\prime
}}^{(p)}\propto \Delta _{j}\Delta _{j^{\prime }}$. To obtain a solution,
Eqs.(\ref{matrix vert eq gamma2}) and (\ref{matrix gap eq}) must first be
left-multiplied by $\boldsymbol{D}^{(p)}=\boldsymbol{W}^{-1}\cdot $.


\begin{thebibliography}{99}
\bibitem{Bogo} N. N. Bogolyubov, V.V. Tolmachev, and D. N. Shirkov, \textit{%
A New Method in the Theory of Superconductivity, }(Consultants Bureau, New
York, 1959).

\bibitem{PWA} P.W. Anderson, \textit{Phys. Rev.} \textbf{110}, 827 (1958); 
\textbf{112}, 1900 (1958).

\bibitem{Tony} A. J. Leggett, \textit{Prog. Theor. Phys}. \textbf{36}, 901
(1966).

\bibitem{optics} V. Guritanu,\textit{\ }A. B. Kuzmenko, D. van der Marel, S.
M. Kazakov, N. D. Zhigadlo, and J. Karpinski, \textit{Phys. Rev. B} \textbf{%
73}, 104509 (2006).

\bibitem{Liu} A.Y. Liu, I. I. Mazin, and J. Kortus, \textit{Phys. Rev. Lett.}
\textbf{87}, 087005 (2001).

\bibitem{bandstructure} P. Szabo%
, P. Samuely, J. Kacmarchik, T. Klein, J. Marcus, D. Fruchart, S. Miraglia,
C. Marcenat, and A. G. M. Jansen., \textit{Phys. Rev. Lett.} \textbf{87},
137005 (2001). M. Iavarone, G. Karapetrov, A. E. Koshelev, W. K. Kwok, G. W.
Crabtree, D. G. Hinks, W. N. Kang, Eun-Mi Choi, Hyun Jung Kim, Hyeong-Jin
Kim and S. I. Lee., \textit{Phys. Rev. Lett.} \textbf{89}, 187002 (2002). S.
Tsuda, T. Yokoya, T. Kiss, Y. Takano, K. Togano, H. Kito, H. Ihara, and S.
Shin., \textit{Phys. Rev. Lett.} \textbf{87}, 177006 (2001). S. Souma, Y.
Machida, T. Sato, T. Takahashi, H. Matsui, S.-C. Wang, H. Ding, A. Kaminski,
J. C Campuzano, S. Sasaki and K. Kadowaki., \textit{Nature}\textbf{\ }%
(London) \textbf{423}, 65 (2003).

\bibitem{Blumbergphonons} A. Mialitsin, B. S. Dennis, N. D. Zhigadlo, J.
Karpinski, and G. Blumberg, \textit{Phys. Rev. B}, \textbf{75}, 020509(R)
(2007).

\bibitem{Blumbergelectrons} G. Blumberg, A. Mialitsin, B. S. Dennis, M. V.
Klein, N. D. Zhigadlo, and J. Karpinski,, \textit{Phys. Rev. Lett.} \textbf{%
99}, 227002 (2007).

\bibitem{KD} M. V. Klein and S. D. Dierker, \textit{Phys. Rev. B} \textbf{29}%
, 4976 (1984).

\bibitem{Kawabata} A. Kawabata, \textit{J. Phys. Soc. Jpn.} \textbf{30}, 68
(1971).

\bibitem{Nambu} Y. Nambu, \textit{Phys. Rev.} \textbf{117}, 648 (1960).

\bibitem{sym} One can prove Eq.(\ref{symmetry}) by iterating Eqs.(\ref{Vert
eq gamma 3}) and (\ref{Vert eq gamma2}), starting with $\Gamma _{k,\gamma
}^{\left( 3\right) }(\omega ,\boldsymbol{q})=\gamma _{k}$ and $\Gamma
_{k,\gamma }^{\left( 2\right) }(\omega ,\boldsymbol{q})=0$. The result will
give an infirite series for $B_{\gamma ^{\prime },\Gamma \gamma }(\omega ,%
\boldsymbol{q})$, each term of which is symmetric in the components of $%
\gamma $ and $\gamma ^{\prime }$.

\bibitem{comment} Note that this assumption rules out the simplifying
product ansatz of the form $D_{j,j^{\prime }}^{(p)}\propto \Delta _{j}\Delta
_{j^{\prime }}$.

\bibitem{bubble} The necessity for including bubble diagrams was mentioned
by Nambu\cite{Nambu} in another context, but was not considered by K-D\cite%
{KD}.

\bibitem{phonons} Abhay Shukla, Matteo Calandra, Matteo D'Astuto, Michele
Lazzeri, Francesco Mauri, Christophe Bellin, Michael Krisch, J. Karpinski,
S. M. Kazakov, J. Jun, D. Daghero, and K. Parlinski., \textit{Phys. Rev.
Lett.} \textbf{90}, 095506 (2007); J. Geerk, R. Schneider, G. Linker, A. G.
Zaitsev, R. Heid, K.-P. Bohnen, and H. V. Loehneysen., \textit{Phys. Rev.
Lett.} \textbf{94}, 227005 (2007).

\bibitem{exception} One exception is the paper by Chubukov \textit{et al. }%
See the next reference.

\bibitem{Andrey} A. V. Chubukov, I. Eremin, and M. M. Korshunov, \textit{%
Phys. Rev. B }\textbf{79}, 220501(R) (2009).
\end{thebibliography}
\end{document}